\documentclass[sigconf]{acmart}

\acmConference[ICSE 2024]{46th International Conference on Software Engineering}{April 2024}{Lisbon, Portugal}

\usepackage{chngcntr}

\usepackage{graphicx}

\usepackage{amsmath}

\usepackage{subcaption}

\usepackage{amsfonts}
\usepackage{algorithmic}
\usepackage{xspace}
\usepackage{soul}
\usepackage{url}
\usepackage{multirow, multicol, makecell}
\usepackage{url}
\usepackage{xspace}
\usepackage{framed}
\usepackage[breakable,skins]{tcolorbox}

\newcolumntype{?}{!{\vrule width 1pt}}

\newcommand{\DMS} {Different Meaning Sentence}
\newcommand{\EMS} {Equivalent Meaning Sentence}

\newcommand{\figlang}[1]{{\em `#1'}}
\newcommand{\sent}[1]{{\em ``#1''}}

\title{Shedding Light on Software Engineering-specific Metaphors and Idioms}

\author{Mia Mohammad Imran}
\affiliation{\institution{Virginia Commonwealth University}
  \city{Richmond, Virginia}
  \country{USA}
}
\email{imranm3@vcu.edu}

\author{Preetha Chatterjee}
\affiliation{\institution{Drexel University}
  \city{Philadelphia, Pennsylvania}
  \country{USA}
}
\email{preetha.chatterjee@drexel.edu}

\author{Kostadin Damevski}
\affiliation{\institution{Virginia Commonwealth University}
  \city{Richmond, Virginia}
  \country{USA}
}
\email{kdamevski@vcu.edu}

\begin{abstract}
Use of figurative language, such as metaphors and idioms, is common in our daily-life communications, and it can also be found in Software Engineering (SE) channels, such as comments on GitHub. Automatically interpreting figurative language is a challenging task, even with modern Large Language Models (LLMs), as it often involves subtle nuances. This is particularly true in the SE domain, where figurative language is frequently used to convey technical concepts, often bearing developer affect (e.g., \figlang{spaghetti code}). Surprisingly, there is a lack of studies on how figurative language in SE communications impacts the performance of automatic tools that focus on understanding developer communications, e.g., bug prioritization, incivility detection. Furthermore, it is an open question to what extent state-of-the-art LLMs interpret figurative expressions in domain-specific communication such as software engineering. To address this gap, we study the prevalence and impact of figurative language in SE communication channels. This study contributes to understanding the role of figurative language in SE, the potential of LLMs in interpreting them, and its impact on automated SE communication analysis. Our results demonstrate the effectiveness of fine-tuning LLMs with figurative language in SE and its potential impact on automated tasks that involve affect. We found that, among three state-of-the-art LLMs, the best improved fine-tuned versions have an average improvement of 6.66\% on a GitHub emotion classification dataset, 7.07\% on a GitHub incivility classification dataset, and 3.71\% on a Bugzilla bug report prioritization dataset.
\end{abstract}

\begin{document}

\maketitle

\begin{sloppypar}

\section{INTRODUCTION}

Figurative language is the use of words or phrases in a way that deviates from their literal meaning, aiming to evoke specific concepts or imagery within one's imagination\footnote{\url{https://www.ldoceonline.com/dictionary/figurative}}. Figurative language consists of different types~\cite{giora1999understanding}, such as metaphors, which use comparisons to describe something differently (e.g., \sent{the road ahead is a long and winding journey}); idioms, which are common phrases that have alternate meanings (e.g., \figlang{to beat around the bush}); similes, which use \figlang{like} or \figlang{as} to compare two things (e.g., \figlang{as light as a feather}); and personification, which gives human qualities to objects or animals (e.g., \figlang{the leaves danced in the wind}).

Within the software engineering (SE) community, professionals often employ various distinctive figurative expressions that are not commonly used in everyday discourse. For instance, developers utilize the metaphorical term \figlang{anti-pattern} to communicate the idea of a recurring problem that should be avoided~\cite{neill2011antipatterns}. Idioms, another frequently employed form of figurative expression, play a crucial role in SE communication by succinctly and colloquially conveying common ideas or concepts. An example of this is when developers describe poorly written code as \figlang{spaghetti code}, implying that it is convoluted and challenging to comprehend~\cite{steele1977macaroni}.

Just as humans use phrases like \figlang{boil} with anger or \figlang{a breath of fresh air} for relief~\cite{kovecses2002emotion}, developers might say \figlang{a thorn in my side}\footnote{\url{https://github.com/chipsalliance/chisel/pull/3352\#issuecomment-1593479230}} to express persistent annoyance or difficulty with an API or a feature. Use of pejorative terms like \figlang{garbage code}~\footnote{\url{https://github.com/adamit24/countdownClass/issues/1}}, \figlang{Frankencode}~\footnote{\url{https://github.com/drupal-code-builder/drupal-code-builder/issues/165}} can be indicators of severe negative emotions, leading to toxic discussions.
Therefore, understanding the use of figurative language in software development discourse can help detect the use of offensive language~\cite{plaza2022integrating} and provide valuable insights into the overall health of a software project~\cite{ferreira2021shut}.
Developers also use figurative expressions to indicate the impact and severity of a bug. For instance, while expressions like \figlang{a ticking time bomb}~\footnote{\url{https://github.com/godotengine/godot/issues/77480\#issue-1726201628}}, suggests significant future problems,  \figlang{showstopper}~\footnote{\url{https://github.com/conwid/VSCleanBin/issues/2\#issuecomment-571094076}} and \figlang{critical roadblock}~\footnote{\url{https://github.com/canjs/canjs/pull/3286}}, emphasize the urgency of addressing the bug at the present.
A recent blog article noted how an improved understanding of software engineering metaphors would mitigate the risks of misinterpretations, fostering more precise and effective communication within the software development community\footnote{\url{https://www.leadingagile.com/2018/11/metaphorically-speakingthe-history-of-communication-in-software/}}, while a recent study showed figurative language like Humor has positive effect on developer engagement~\cite{tiwari2023great}.
Recent research studies have also highlighted that flaws in SE emotion and sentiment detection tools stem from the use of figurative language~\cite{novielli2021assessment, novielli2020can, imran2022data, chen2021emoji}. On a Stack Overflow and GitHub dataset, Novielli et al.~\cite{novielli2021assessment} found that 9\% of the errors in sentiment analysis were due to figurative language, noting that it poses an open challenge for sentiment detection in software engineering.

Despite its potential impact on the performance of automatic tools focused on understanding SE text, there have been very limited studies on analyzing figurative language in SE, e.g., there have been some studies on SE synonyms~\cite{8827962,chen2017unsupervised} and programming language-specific idioms~\cite{alexandru2018usage}.
In this paper, we aim to go beyond the synonyms and explore the broader landscape of figurative language in SE. We aim to \figlang{shed light on} or analyze the use of figurative language (specifically, metaphors and idioms) in SE communication channels and contribute to the understanding of how recently proposed language models that target software-related text can be made to recognize figurative expressions.

Large Language Models (LLMs), such as BERT~\cite{bert} and RoBERTa~\cite{roberta}, have recently demonstrated state-of-the-art results on a variety of software engineering tasks, e.g., code completion, code review, bug localization, sentiment analysis, toxicity detection~\cite{ciborowska2022fast, ciniselli2021empirical, linsentiment, sghaier2023multi, zhang2020sentiment, sarker2020benchmark}. While LLMs are not explicitly designed to detect figurative languages like metaphors and idioms, they can acquire this ability through training on large datasets such as Wikipedia and Stack Overflow~\cite{su2020deepmet, gamage2022bert, BRISKILAL2022102756, song2021verb}. This capability is particularly beneficial in the software engineering context, as it enables a more nuanced and accurate analysis of developer communications. Without this ability, an LLM may misinterpret or misclassify text, leading to erroneous results. For instance, if an LLM cannot recognize the idiom \figlang{edge case}, it may interpret the phrase literally and erroneously categorize the text as being related to a specific type of physical boundary instead of grasping its figurative meaning of a rare or unusual scenario.

Through this study, we will examine the relevance of figurative language in GitHub communication channels, the ability of LLMs to detect figurative language in the SE context, and the impact of figurative language on affect analysis and bug report priority detection. By gaining a deeper understanding of the role and effects of figurative language in SE, we aim to contribute to the development of more effective and accurate NLP-based systems for SE tasks, specifically in automated recognition of developer emotions and incivility on GitHub, and bug report priority detection. We focus on answering the following three research questions:

\smallskip
\noindent
{\em RQ1: How well can existing LLMs interpret figurative language (i.e., metaphors and idioms) used in software engineering?}

\noindent
To answer this RQ, we collect a set of 2000 sentences containing figurative language and create \textit{rephrased} sentences, i.e., sentences with similar meanings but without figurative expressions. We also create \textit{altered} sentences that share as many words as the original sentence but convey different meanings, e.g., using metaphors in their literal sense or using idioms in a different context other than software engineering.
This procedure of creating, so called, entailed and non-entailed text from premise text has been widely used in NLP~\cite{stowe2022impli, bowman2015large, bowman2020new}. Using this data triple of original, rephrased, and altered meaning sentences,
we investigate whether LLMs can recognize the semantics of figurative sentences by computing how often the models identified the semantic dissimilarity of the rephrased sentence with the altered sentence. Our results suggest that LLMs have a limited ability to interpret figurative language, with higher performance for general figurative expressions than software engineering-specific ones.

\smallskip
\noindent
{\em RQ2: Can the performance of software engineering-specific affective analysis be improved by a better insight into figurative language?}

\noindent
Affect expressions are the means to convey emotions, feelings, and attitudes to others~\cite{mohammad2016sentiment}. For some time now, researchers have been exploring automatic affect analysis, which encompasses tasks such as emotion analysis, sentiment analysis, and incivility analysis. To answer RQ2, we fine-tune several LLMs using contrastive learning~\cite{contrastive} with our dataset of figurative language in order to improve their ability to interpret figurative language. We then compare the performance of the fine-tuned LLMs to the original models of two publicly available affect datasets: an emotion dataset curated from GitHub, and an incivility dataset curated from GitHub. Our results indicate that fine-tuned LLMs perform better in both cases.

\smallskip
\noindent
{\em RQ3: Can a better understanding of figurative language enhance software engineering automation where affect plays a role?}

\noindent
A number of research tasks in SE indirectly involve affective natural language text, e.g., app review analysis, opinion mining~\cite{Chatterjee2021AutomaticEO, lin2022opinion}. Specifically, in this RQ we investigate how a better understanding of figurative language can impact bug report priority detection, which is a significant area of interest in open-source software research~\cite{wang2022clebpi, tian2015automated, umer2018emotion, umer2019cnn}. Umer et al. observed that emotions influence bug report priority detection~\cite{umer2018emotion}. To address this problem, recently researchers have employed Language Models (LLMs)~\cite{wang2022clebpi}. In this study, we explore LLMs fine-tuned with contrastive learning using our figurative language dataset, similar to the approach in RQ2, and conducted experiments on the publicly available Bugzilla dataset\footnote{\url{https://bugs.eclipse.org/bugs/}}.
Our results indicate that fine-tuning with our figurative language dataset improves bug report priority detection.

We publish the annotation instructions, annotated dataset, and source code to facilitate the replication of our study at \color{blue}\url{https://github.com/vcu-swim-lab/SE-Figurative-Language}\color{black}.
 \section{DATASET}

To conduct our study, we curate a dataset of developer communications containing figurative language. Towards that goal, we first collect data from GitHub issues and pull requests and identify the occurrences of idioms and metaphors. To inquire whether language models understand figurative language, we manually rephrase the original sentences containing figurative language to generate: 1) sentences that are similar in meaning to the original but do not contain idioms or metaphors; and 2) sentences that contain similar words as the original sentences but are semantically dissimilar, i.e., have a different meaning.
In this section, we detail each step involved in constructing our dataset.

\subsection{Data Collection}

We selected nine popular GitHub repositories, each with a minimum of 50k stars: {\sf skylot/jadx}, {\sf laravel/laravel}, {\sf microsoft/PowerToys}, {\sf rails/rails}, {\sf redis/redis}, {\sf facebook/react}, {\sf tensorflow/tensorflow}, {\sf huggingface/transformers}, and {\sf microsoft/vscode}. We collected 10k comments from each repository (5k PR comments and 5k issue comments) between February 2022 and May 2023.
We split the comments into sentences using NLTK\footnote{\url{https://www.nltk.org/}} and filtered out sentences with fewer than 5 words, resulting in a total of 202k sentences.

One of our study's end goals is to examine figurative language's impact on affective expressions in a software engineering context. Previous research has shown that most comments on GitHub are neutral, lacking any detectable emotions or sentiments~\cite{Murgia2014DoDF}. Therefore, we excluded neutral sentences by using a software engineering-specific sentiment analysis tool~\cite{eeshita-sentiment}.

In addition, to avoid including sentences that do not contain any figurative expressions, we applied a popular metaphor detection~\cite{su2020deepmet} and an idiom detection tool~\cite{gamage2022bert} to identify candidate metaphors and idioms in each sentence. This model-in-the-loop approach is popular in Natural Language Inference (NLI) research, e.g., figurative language interpretation~\cite{nie2019adversarial, chakrabarty2022flute}, as it maximizes the value of annotation effort, which requires tedious human labor. We discarded sentences that do not contain any candidate idioms or metaphors. We randomly selected 1000 sentences containing metaphors from the remaining sentences. We also randomly chose 1000 sentences containing idioms (different from the metaphor set). This process resulted in a dataset of 2000 sentences.

\subsection{Data Annotation}\label{data}

\begin{figure}[t]
\centering
\includegraphics[width=\linewidth]{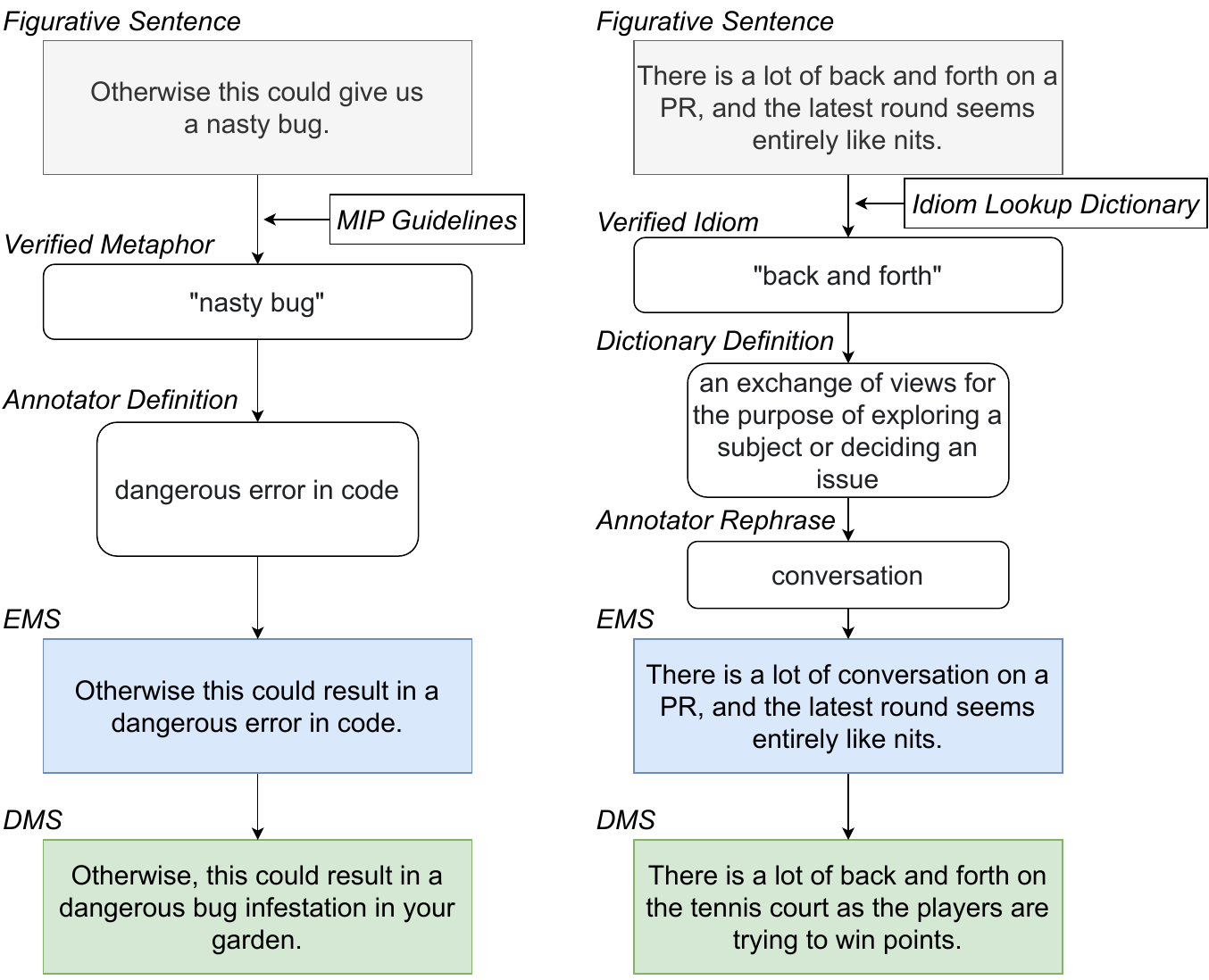}
\caption{Figurative language annotation procedure.}
\label{fig:annotation}
\end{figure}

First, we recruited four annotators (two graduate students and two senior undergraduate students) who were each given 500 sentences to annotate (250 with metaphors and 250 with idioms). Due to the nature of the task and difficulties with crowd-sourcing~\cite{bowman2020new}, we opted for a small number of annotators
that are native speakers/professionally fluent in English with a strong computer science background.
Along with the 2000 sentences in total, the annotators were provided with a set of candidate figurative expressions marked by the above-mentioned tools. We instructed them to: 1) verify the candidates as metaphors or idioms and judge whether each metaphor or idiom is specific to software engineering or general purpose; and 2) create rephrased sentences from the original. We also held a short training session in which we reviewed the annotation process for a few representative examples with each annotator. Below, we describe these data annotation steps in detail (see also Figure~\ref{fig:annotation}).

\subsubsection{Verifying Figurative Expressions} \label{verified} For verifying metaphors and idioms, we followed best practices from existing literature. More specifically, to verify the metaphors we asked the annotators to carefully read the Metaphor Identification Procedure (MIP) guideline by the Pragglejaz Group~\cite{group2007mip}. The MIP guideline is a well-known procedure for identifying metaphors. Based on the guideline, the annotators marked the correct metaphoric expressions from the candidate set. For example, the annotators confirmed that \figlang{nasty bug} is a valid metaphor for a difficult fault in the sentence, \sent{Otherwise, this could give us a nasty bug.}

We noted in the annotation instructions that most metaphors are conventional, i.e., metaphors that are often used in everyday language~\cite{do2018weeding}. For example, in the following sentence: \sent{I see your point}, \figlang{see} and \figlang{point} both are metaphors~\cite{do2018weeding}. Often such cases can be observed in software engineering communication. For instance, \figlang{pinging} in the following sentence is a metaphor: \sent{Hi @[USER], thanks for pinging me on this issue.} Here, \figlang{pinging} is a colloquial way of saying \figlang{contacting someone}, while the literal meaning of \figlang{pinging} comes from computer networking terminology\footnote{\url{https://ftp.arl.army.mil/~mike/ping.html}}.

For verifying idioms, we followed the guideline provided by Stowe et al.~\cite{stowe2022impli}, which asked the annotators to look up idioms in popular dictionaries (such as the Oxford English Dictionary\footnote{\url{https://www.oxfordlearnersdictionaries.com/}}, the Webster Dictionary\footnote{\url{https://www.merriam-webster.com/}}, and the Longman Dictionary of Contemporary English\footnote{\url{https://www.ldoceonline.com/}} and popular search engines (e.g., Google).
We instructed the annotators to consider an expression as likely to be an idiom if its dictionary definition is: 1) applicable in the context; and 2) a good syntactic fit in the same environment. For example, in the sentence, \sent{I will also be keeping an eye on you}, \figlang{keeping an eye} is an idiom which means \figlang{to watch someone or something or stay informed about the person's behavior, especially to keep someone out of trouble.}\footnote{\url{https://dictionary.cambridge.org/}} Conversely, when the meaning of the candidate idiomatic expression is literal in the context of the sentence and the dictionary definition is not applicable, it is likely not to be an idiom. For example, in the sentence, \sent{It was cold, so cold in the jeep that it was with difficulty that Alexei kept his eyes open}, \figlang{kept his eyes open} is not an idiom.
Since software-specific words have distinct meanings from conventional terms (e.g., bug, issue, error, function), we supplied annotators with established software engineering glossary terms from the FDA\footnote{\url{https://www.fda.gov/inspections-compliance-enforcement-and-criminal-investigations/inspection-guides/glossary-computer-system-software-development-terminology-895}} and Google\footnote{\url{https://developers.google.com/machine-learning/glossary}}.

Once annotators verified the candidate set, we asked them to mark whether the figurative expressions were software engineering-specific or general-purpose. The annotators identified 752 sentences with metaphors and 909 with idioms, totaling 1661 sentences. The remaining 339 sentences did not contain any figurative words. These 1661 sentences had a total of 1741 unique figurative expressions, with 445 being SE-specific and 1296 general.

\subsubsection{Rephrasing Sentences}
The process of rephrasing sentences was divided into two phases: creating semantically-equivalent rephrased sentences and constructing altered-meaning sentences. We refer to the semantically equivalent rephrased sentences as {\EMS} (EMS) throughout the paper. These sentences retain the original meaning of the sentence, but the figurative expressions are replaced with literal terms. We refer to the altered-meaning sentences as  {\DMS} (DMS). These sentences are modified so that they significantly differ in meaning from the original sentences.

\smallskip
\noindent
\textit{a) EMS Construction}: The annotators were tasked with rephrasing each sentence on their list, i.e., removing the (verified) figurative expressions while maintaining the original semantics of the sentence as much as possible. In other words, the replaced figurative expression should entail its literal counterpart. For example, in the sentence, \sent{[USER] Thanks for your help, what you said may be a hidden bug.}, the figurative expression \figlang{hidden bug} is replaced with \figlang{unseen error} resulting in the {EMS}: \sent{[USER] Thanks for your help, what you said may be an unseen error.} This approach is inspired by previous research by Stowe et al. on figurative language in NLP~\cite{stowe2022impli}. It is worth noting that for EMS we did not employ multiple annotators to annotate the same set or calculate inter-annotator agreement as Stowe et al. found that this method does not yield significantly different quality compared to the conventional approach~\cite{stowe2022impli}.

\smallskip
\noindent
\textit{b) DMS Construction}: {\DMS}s are variations of metaphorical or idiomatic sentences that convey a different meaning than the original sentence and do not entail it~\cite{stowe2022impli}. Two strategies were employed to construct DMS: a) using figurative expressions in a literal sense; and b) replacing the figurative expressions and their context with different words. These strategies are inspired by previous research on figurative language in natural language processing~\cite{stowe2022impli, zhou2021pie, haagsma2020magpie}. We also apply a model-in-the-loop approach for DMS generation~\cite{nie2019adversarial, chakrabarty2022flute}. More specifically, we first generated four candidate DMSs for each sentence, two each for each strategy using GPT-4~\cite{OpenAI2023GPT4TR} API (\sf `gpt-4'\footnote{\url{https://platform.openai.com/docs/models/gpt-4-and-gpt-4-turbo/}}), and, second, we recruited human annotators to select the best-generated candidate (or to create one of their own if none is available).

\smallskip
\noindent
\textul{Candidate DMS Generation.} ChatGPT~\cite{OpenAI2023GPT4TR} has shown promising results in data annotation tasks, including text generation, in some cases outperforming human crowd-workers~\cite{gilardi2023chatgpt, huang2023chatgpt}. Following recent literature, we create two GPT-4 prompts for the two different strategies for DMS generation~\cite{kocon2023chatgpt, huang2023chatgpt}. The prompts were carefully devised by using the existing literature on this topic~\cite{stowe2022impli, zhou2021pie, haagsma2020magpie}.
The prompts for generating DMS are as follows:

\smallskip
\noindent
\textit{Generating DMS by using the figurative language in a literal manner:}
\begin{tcolorbox}[breakable,colback=black!10!white]
{\sf \noindent
You are reading GitHub comments with figurative expressions. Your task is to generate 2 examples by using the given figurative expressions in a literal manner to construct different sentences. Do not replace them. Add/change new contexts if necessary. The new sentence must have a completely different meaning than the original. You must keep the semantic order of the original sentences as much as possible. Don't explain your answer.\\\\
Original Sentence:\textcolor{blue}{\em <insert utterance>}.\\\\
Figurative expressions: \textcolor{blue}{\em <insert figurative expressions>}
}
\end{tcolorbox}

\smallskip
\noindent
\textit{Generating DMS by replacing the figurative language:}
\begin{tcolorbox}[breakable,colback=black!10!white]
{\sf \noindent
You are reading GitHub comments with figurative expressions. Your task is to generate 2 examples by replacing given figurative expressions to construct different sentences. The new sentence must have a completely different meaning than the original. You are only allowed to change the figurative expression and its context. You must keep the semantic order of the original sentences as much as possible. Don't explain your answer.\\\\
Original Sentence:\textcolor{blue}{\em <insert utterance>}.\\\\
Figurative expressions: \textcolor{blue}{\em <insert figuration expressions>}
}
\end{tcolorbox}

\smallskip
\noindent
\textul{DMS Selection.} Two additional annotators (one of the authors and one senior undergraduate student) were responsible for the candidate selection of the DMS. We provided the annotators the original sentence, the figurative expressions, and the list of candidate DMSs with the following instructions: \sent{You will be provided with 4 candidate sentences, two of which come from Type 1 and two come from Type 2. Choose the best 1 out of the 4 candidates, with a preference towards choosing from Type 1. If none of these 4 are good candidates, write None. When choosing, try to choose a sentence that has 1) similar semantic order to the original sentence, and 2) a different meaning than the original sentence.}

We instructed the annotators to write their own DMS when their selection is \figlang{None}.
Once they completed an annotation pass over the entire dataset, the two annotators met in person in order to discuss the 310 cases where they disagreed (i.e., selected different DMS candidates or \figlang{None}) and resolved them in order to achieve 100\% agreement.
This human-in-the-loop methodology helps with the more difficult task of DMS generation, enhancing the overall quality and efficiency the process. The iterative resolution of differences ensured a high quality of annotated data.  \section{PREVALENCE OF SE-SPECIFIC FIGURATIVE LANGUAGE}

In order to understand if SE-specific figurative language appears frequently in the wild, we examine the frequency of occurrence of figurative language in a large sample of developer communication on GitHub. More specifically, we collected 1,000 issue comments and 1,000 pull request comments for each of the top 100 repositories by star count on GitHub, i.e., a total of 200k comments. We analyzed comments made from September 1, 2022, to January 1, 2023, spanning 4 months, and excluded repositories with fewer than 1,000 issue and pull request comments during this time.
The collected 200k comments were split into a total of 484k sentences using NLTK\footnote{\url{https://www.nltk.org/}}. Leveraging our annotated dataset consisting of 1741 unique figurative expressions (445 SE-specific and 1296 general), we searched for matches in the set of 484k sentences after applying standard NLP pre-processing (removing punctuation and non-alphabet characters, and lemmatization using SpaCy) since some of the figurative expressions have different spelling variations (e.g., \figlang{root cause}, \figlang{root-cause}, and \figlang{root causes}). To ensure that the matches were not spuriously identifying figurative language (due to polysemy), we also executed the metaphor and idiom detection tools~\cite{su2020deepmet, gamage2022bert}, the same ones that we use for candidates generation, selecting only the matches that were also confirmed by one of these tools.

Of the examined 484k sentences, the 445 SE-specific figurative expressions that annotators identified occurred in 44k sentences (9\%), while the 1296 general figurative expressions occurred in 107k sentences (22\%). Some sentences (2.67\%) had both SE-specific and general figurative expressions.

\begin{figure}[t]
\centering
\begin{subfigure}[t]{0.48\textwidth}
    \centering
    \includegraphics[width=\linewidth]{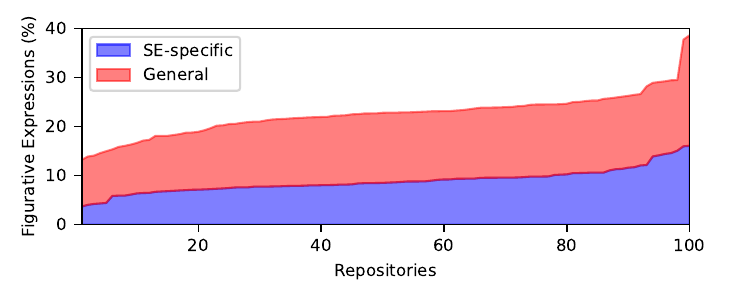}
    \caption{Percent of sentences with figurative expressions in repos.}
    \label{fig:figurative_expressions_percentage}
\end{subfigure}
\begin{subfigure}[t]{0.48\textwidth}
    \centering
    \includegraphics[width=\linewidth]{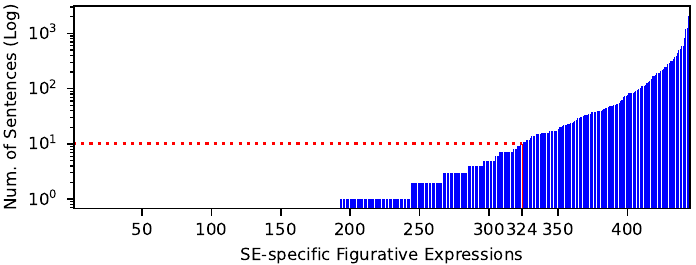}
    \caption{Frequency of sentences with SE-specific figurative expressions.}
    \label{fig:se_specific_counter}
\end{subfigure}
\hfill
\caption{Distribution of figurative language occurrence in GitHub sentences (200k GitHub comments, 484k sentences).}
\label{fig:top_repos}
\end{figure}

The distribution of general and SE-specific figurative expressions in different GitHub repositories is shown in Figure~\ref{fig:figurative_expressions_percentage}. SE-specific figurative language does appear in non-trivial amounts in most repositories we examined (i.e., in between 3.69\% and 16.08\% of sentences), but much less often than general figurative expressions, which occurred in between 13.2\% and 38.62\% of sentences.
In Figure~\ref{fig:se_specific_counter}, we present the frequency of SE-specific figurative expressions identified within our corpus of 200k GitHub comments. Of the 445 SE-specific figurative expressions we investigated, 324 (72.8\%) appear no more than 10 times, as indicated by the red dotted line, suggesting that most such expressions are infrequent. Among these, 193 expressions are absent from our dataset. This absence can be attributed to two main factors: a) expressions that are specific to particular projects (e.g., \textit{`ghost highlight'} and \textit{`ghost monitor'} in UI-related projects); and b) unique expressions used to describe highly specific scenarios (e.g., \textit{`dead fork,'} \textit{`magic code'}).

Note that our study provides only a lower bound, as it matches using an incomplete set of figurative language expressions. Therefore, the likely presence of figurative language is even higher than we report. This exploratory study highlights the importance of understanding figurative language in the SE context, as it can provide insight into the daily communications of developers.

 \section{EXPERIMENTS AND DISCUSSION}
Using the assembled dataset, we created specific experiments for each of our three research questions. In this section, we describe the experiments and discuss the corresponding results.

\begin{table*}[t]
\centering
\footnotesize
\caption{Percent of EMS with a higher similarity to the original sentence than corresponding DMS (Sim$_{EMS}$ > Sim$_{DMS}$).}
\begin{tabular} { |l|c|c|c|c|c|c|c|c|c|} \hline
    Model & \multicolumn{3}{c|}{\textit{SE-specific}} & \multicolumn{3}{c|}{\textit{General}} & \multicolumn{3}{c|}{\textit{Overall}} \\ \cline{2-10}
    & Sim$_{EMS}$ > Sim$_{DMS}$ & {\em p-value} & |Cliff's $\delta$| & Sim$_{EMS}$ > Sim$_{DMS}$ & {\em p-value} & |Cliff's $\delta$| & Sim$_{EMS}$ > Sim$_{DMS}$ & {\em p-value} & |Cliff's $\delta$|
    \\ \hline \hline
    BERT & 84.51\%  &  $p$ < 0.01 & 0.629 & 87.40\%  &  $p$ < 0.01 & 0.638 & 86.57\% &  $p$ < 0.01 & 0.637 \\ \hline
    RoBERTa & 83.70\% & $p$ < 0.01 & 0.648 & 85.21\% & $p$ < 0.01  & 0.620 & 84.95\% & $p$ < 0.01  & 0.632 \\ \hline
    ALBERT & 81.79\% & $p$ < 0.01 & 0.610 & 85.80\% & $p$ < 0.01  & 0.598 & 85.00\% & $p$ < 0.01  & 0.605 \\ \hline
    CodeBERT & 77.99\% & $p$ < 0.01 & 0.498 & 79.63\% & $p$ < 0.01  & 0.493 & 79.11\% & $p$ < 0.01  & 0.495 \\ \hline
\end{tabular}
\label{tab:cosine_similarity}
\end{table*}

\subsection{RQ1: How well can existing LLMs interpret figurative language (i.e., metaphors and idioms) used in software engineering?}

In order to determine how well popular Large Language Models (LLMs) understand metaphors and idioms, we examine whether they can understand the semantic relationship between the original sentence, the equivalent sentence (i.e., EMS), and the different-meaning sentence (i.e., DMS). The task of differentiating EMSs from DMSs of the original sentences can be thought of as Recognizing Textual Entailment (RTE)~\cite{chakrabarty2021figurative}. RTE involves determining whether a statement, called the hypothesis, can be inferred from a given text, called the premise. In our context, the premise is the original sentence, and the hypotheses are the EMS and DMS. We evaluate whether an LLM can infer the EMS from the original sentence, and if it is a DMS, the model should not deduce it.

One way to approximate the RTE task is to posit that the model should recognize the original sentence as semantically closer to the EMS than to the DMS. By comparing the embedding vectors of the original sentence and the sentence in question, we can measure whether the two sentences are similar or dissimilar and, therefore, whether the most similar sentence is an EMS or a DMS.

\subsubsection{Compared LLMs} We compare three LLMs -- BERT~\cite{bert}, RoBERTa~\cite{roberta}, and ALBERT~\cite{albert}, which are popular in NLP and SE tasks~\cite{batra2021bert, eeshita-sentiment, mao2022biases}. BERT~\cite{bert} is a transformer-based model pre-trained on extensive text data from Wikipedia and BooksCorpus.
RoBERTa~\cite{roberta} is an improved version of BERT, while ALBERT~\cite{albert} enhances efficiency through parameter reduction techniques.
Additionally, we evaluate a popular SE-specific LLM -- CodeBERT~\cite{codebert}, which is pre-trained on natural language - programming language pairs. We use {\em bert-base-uncased}, {\em roberta-base}, {\em albert-base-v2} and {\em microsoft/codebert-base} available from Hugging Face\footnote{\url{https://huggingface.co}}.

\begin{figure}[tb]
\centering
\includegraphics[width=0.82\linewidth]{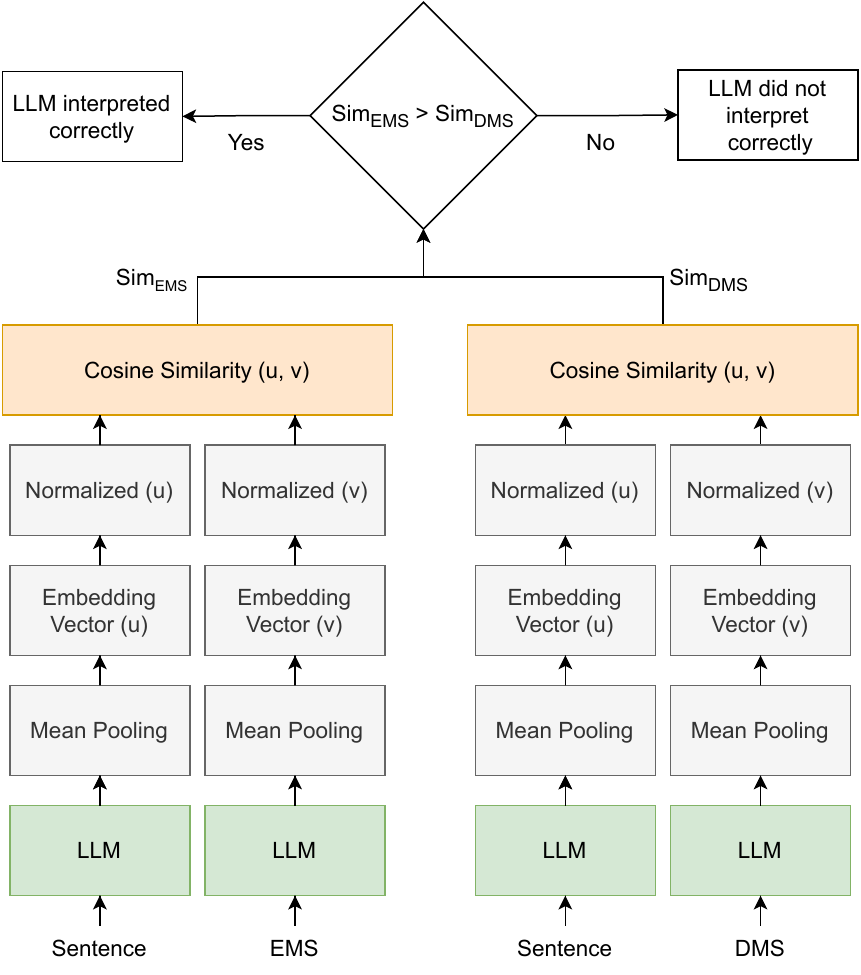}
\caption{RQ1 evaluation pipeline.}
\label{fig:rq1_methodology}
\end{figure}

\subsubsection{Procedure}

We generate embedding vectors using each of the LLMs for each pair of sentences, i.e., {\em <original sentence, EMS>} and {\em <original sentence, DMS>}. Next, we compute the similarity between the vectors in each pair and then compare the resulting similarity scores~\cite{reimers2019sentence}. We perform standard software engineering text-specific preprocessing operations such as URL removal, username removal, stack trace removal, etc.~\cite{imran2021automatically}

Since the LLMs in our cohort produce word-level embedding vectors, there are several possibilities for aggregating these into sentence-level embedding vectors. Reimers et al.~\cite{reimers2019sentence} noted that mean pooling (the mean of all per-word output vectors generated by the LLM) is one of the best strategies. While we opt for mean pooling when generating each sentence's embedding vector, we also note that this strategy is still error-prone due to the anisotropy problem, i.e., the difference in the scale of the embedding vectors~\cite{ethayarajh2019contextual}.
For this reason, we apply the normalization proposed by Yan et al.~\cite{yan2022addressing}, which is based on Singular Value Transformation (SVT). SVT  uses singular value decomposition and a threshold using the soft-exponential function by Godfrey et al.~\cite{godfrey2015continuum}

Following normalization, we compute vector pair similarity with the cosine similarity metric.
Cosine similarity measures vector alignment, with values from 1 (identical) to 0 (orthogonal) to -1 (opposite)~\cite{gomaa2013survey}.
Then, we compare the two similarities in order to determine if the {\em <original sentence, EMS>} similarity (Sim$_{EMS}$) is higher than the {\em <original sentence, DMS>} similarity (Sim$_{DMS}$). Figure~\ref{fig:rq1_methodology} summarises the entire procedure.
To evaluate the RQ, we compute the percentage of instances where Sim$_{EMS}$ is greater than Sim$_{DMS}$. We examine three sentence categories: a) those containing SE-specific figurative language only (n=371); b) those containing general figurative language only (n=1179); and c) overall, containing either (a) and (b), or both (n=1661).
For each model and category, we measure the statistical significance of the difference between the two cosine similarities using the one-tailed Wilcoxon signed-rank test (i.e., testing if Sim$_{EMS}$ > Sim$_{DMS}$ with statistical significance). We apply the Benjamini-Hochberg correction to control the false discovery rate. A small {\em p-value} (e.g., {\em p-value} < 0.05) indicates that the difference is unlikely to be due to chance and that there is a statistically significant difference between the two samples. We also compute the effect size, which measures the magnitude of the difference between the two samples, using Cliff's Delta ($\delta$)~\cite{cliff1993dominance}, where |$\delta$| > 0.147, 0.33, and 0.474 indicate small, medium, and large effects respectively.

\subsubsection{Results and Discussion} Table~\ref{tab:cosine_similarity} shows the SE-specific, General, and Overall (i.e., combined) results. The higher the percentage of <\textit{original sentence}, \textit{EMS}> pairs with larger cosine similarity, i.e., Sim$_{EMS}$ > Sim$_{DMS}$, the better the model is at recognizing figurative language. The results table shows that the BERT and RoBERTa models have the highest percentage of correctly understood pairs for all categories. BERT, RoBERTa, and ALBERT models correctly recognize 84.51\%, 83.70\%, and 81.79\% of sentences containing SE-specific figurative expressions, 87.40\%, 85.21\%, and 85.80\% of General figurative expressions, and 86.57\%, 84.95\%, and 85.0\% of the Overall figurative expressions, respectively.
In the case of CodeBERT, which exhibits the poorest results out of all models, there is no significant difference between SE-specific and General results (77.99\% and 79.63\% respectively). This is likely because the model is pre-trained with programming-specific data enabling it recognize some software engineering figurative language terms. However, it also likely loses the ability to capture General figurative language, which is present in the other LLMs.
From this study, we observe that all of the models can understand figurative language to a reasonable degree (i.e., ranging between 77.99\% to 87.40\%).

This is also evident from {\em p-value} and Cliff's |$\delta$|. In each case, the statistically significant is with a {\em p-value} < 0.01 and a |$\delta$| greater than 0.474 is considered a large effect for all models.
The p-value less than 0.01 indicates that the observed difference in similarity between the two groups is highly unlikely to be due to chance and we conclude that there is a statistically significant difference in similarity between the two sets of sentence pairs.
The large |$\delta$| indicates that the similarity between the two groups (i.e., the sentence pairs in group Sim$_{EMS}$ compared to those in group Sim$_{DMS}$ is substantial. The cosine similarity values in Group Sim$_{EMS}$ are consistently higher than those in Group Sim$_{DMS}$, showing that the sentences in Group Sim$_{EMS}$ are more similar to each other compared to those in Group Sim$_{DMS}$.
Together, these results suggest that the two groups of sentence pairs exhibit a notable and meaningful difference in their similarity scores, and this difference is not likely to be due to random chance.

However, the models still fail to recognize between 18.21\% and 12.60\% of figurative language instances. It is highly likely that if we can improve the models' understanding of figurative language in such cases, they will function better in their use cases.

\subsection{RQ2: Can the performance of software engineering-specific affective analysis be improved by a better insight into figurative language?}

Affect analysis involves identifying and evaluating human emotions, feelings, and sentiments expressed through written communication.
Kovecses et al. noted that figurative expressions are vital in expressing emotions ~\cite{kovecses2002emotion}, while Mohammad et al.~\cite{mohammad2016metaphor} observed that metaphorical words tend to contain significantly more emotions than the literal sense of the same words.
In software engineering, affect is often related to the software and its development process, including the emotional states of software developers, productivity, and burnout~\cite{graziotin2015feelings, raman2020stress, graziotin2017unhappiness}. Thus, identifying and understanding affect is crucial for improving software quality and developer productivity. However, several studies have shown challenges in building reliable tools and datasets for mining emotions and opinions in the SE domain~\cite{novielli2018benchmark, lin2022opinion}. A recent study by Imran et al. found that using figurative language in SE-related text can hinder the accurate identification of emotions ~\cite{imran2022data}, partly motivating our RQ2 investigation.

The use of LLMs has become a widely adopted method for identifying and classifying affective expressions in written text~\cite{sun2019fine, ke2021adapting, chiorrini2021emotion}. LLMs are usually fine-tuned to address specific affect analysis tasks, such as recognizing sentiment or emotions. Recently, one of the most effective ways to fine-tune an LLM is by applying a contrastive learning approach. This approach uses sets of similar and dissimilar instances to train the model to understand the similar instances and differentiate them from the dissimilar ones~\cite{liu2022testing}. To answer this RQ, we leverage contrastive learning as the means for LLMs to better capture the meaning and nuances in figurative language present in GitHub comments.

\subsubsection{Compared models} Similar to RQ1, we assess the ability of the same four LLMs --- BERT~\cite{bert}, RoBERTa~\cite{roberta}, ALBERT~\cite{albert}, CodeBERT~\cite{codebert} --- with the same model versions as RQ1 from Hugging Face. Previous research shows that BERT, RoBERTa and ALBERT work well in SE affect analysis~\cite{zhang2020sentiment, batra2021bert, wang2022clebpi}.

\subsubsection{Contrastive learning} Contrastive learning is a recently proposed machine learning technique that involves training a model
to distinguish between two or more distinct data points by contrasting their differences~\cite{contrastive, contrastive_2}.
The steps for applying this approach to fine-tune LLMs for understanding figurative language elements in the text can be outlined as follows:

\begin{enumerate}
    \item The LLM is presented with a triplet of anchor, positive and negative samples, which are representative of the figurative language elements to be learned.
    \item The LLM processes the samples and generates output embeddings for the data triplet.
    \item A loss function encourages the anchor and positive samples to be closer together and the anchor and negative samples to be further apart in the embedding space.
    \item The process is repeated until the LLM has learned a satisfactory representation.
\end{enumerate}

To apply contrastive learning, we use the original sentences and EMSs as anchor and positive classes and DMSs as negative classes. In other words, we created {\em <original sentence, EMS, DMS>} and {\em <EMS, original sentence, DMS>} as a pair of triplets, where the first element in each pair is anchor, the second element is positive, and the third element is negative. There is a total of 3322 such triplets of sentences in our dataset.

We use InfoNCE Loss as our loss function~\cite{contrastive}. Given the embeddings of an anchor, a positive, and a negative sample denoted as $a$, $p$, and $n$ respectively, the InfoNCE loss is computed as follows:
$
    \text{InfoNCE Loss}(a, p, n) = -\log\left(\frac{e^{\text{sim}(a, p)}}{e^{\text{sim}(a, p)} + e^{\text{sim}(a, n)}}\right)
$
where $\text{sim}(a, p)$ represents the cosine similarity between the embeddings of the anchor and positive samples, and $\text{sim}(a, n)$ represents the cosine similarity between the embeddings of the anchor and negative samples.
The InfoNCE loss maximizes the log-likelihood of anchor-positive similarity and minimizes anchor-negative similarity. We use the Adam optimizer.

Using contrastive learning, the LLM learns to create embeddings that capture the semantic similarity between the original and EMS while recognizing the semantic differences between the original and DMS.
This allows the LLM to learn a representation that separates the positive and negative samples as much as possible. In this case, the LLM learns to recognize the figurative language elements.

After fine-tuning the models with contrastive learning, we assess their performance in two tasks: emotion recognition and incivility detection.
We compare the performance of these fine-tuned models against baseline models that are not fine-tuned with figurative language.

\subsubsection{Datasets} We apply the LLMs to two SE-affect datasets.

\smallskip
\noindent
\textul{Emotion Dataset.} Imran et al.~\cite{imran2022data} curated a multi-label emotion dataset that is crawled from GitHub. The dataset consists of 2000 data points and six emotion classes: Anger, Love, Fear, Joy, Sadness, and Surprise. The dataset contains 340 (17.0\%) Anger comments, 220 (11.0\%) Love comments, 198 (9.9\%) Fear comments, 422 (21.1\%) Joy comments, 274 (13.7\%) Sadness comments, and 328 (16.4\%) Surprise comments. The rest of the comments are neutral.

\smallskip
\noindent
\textul{Incivility Dataset.} Ferreira et al.~\cite{ferreira2022heated} curated a dataset from GitHub's heated issues for incivility detection. The dataset has three parts: comment level, issue level, and sentence level. We consider only the comment-level dataset in our study, which has three classes: Civil, Uncivil, and Technical. We consider only Civil and Uncivil comments as we are only interested in affective analysis for this RQ. The filtered dataset contains 718 comments, of which 232 (32.3\%) are Civil comments, and 486 (67.7\%) Uncivil comments.

\subsubsection{Procedure and Metrics.}

Using random stratified sampling for each class, we divide all two datasets into train (80\%) and test (20\%) sets~\cite{botev2017variance}. For each task (i.e., incivility detection and emotion detection), we train (or fine-tune) both the LLMs' contrastive learning and baseline versions. In other words, the contrastive learning models are fine-tuned twice, first with contrastive learning and figurative language and second with a task-specific dataset. The baselines are only fine-tuned with the task-specific dataset.

We choose a metric that is frequently used to evaluate classification tasks: {\em F1-score}, which aggregates {\em Precision} and {\em Recall}. Precision is the ratio of true positive instances to the total predicted positive instances, and Recall is the ratio of true positive instances to all instances in the positive class. {\em F1-score} is the harmonic mean of {\em Precision} and {\em Recall}:
$\text{\em F1-score} = 2 * \frac{Precision * Recall}{Precision + Recall}$. We also calculate the micro average version for averaging the F1-score across the classes following previous research~\cite{imran2022data, suhaimi2022comparative}.

\subsubsection{Results and Discussion} \textbf{\textit{Emotion Classification:}} Table~\ref{tab:classification_results_emotions} shows the results of the emotion classification task on Imran et al.'s multi-label emotion dataset~\cite{imran2022data}, using BERT, RoBERTa (RBTa), ALBERT (ALBT), CodeBERT (CodBT) and their fine-tuned with figurative language counterparts (BERT-FL, RBTa-FL, ALBT-FL, CodBT-FL). The table presents the F1-score for each emotion class, the micro-averaged F1-score, and the improvement in the F1-score achieved by the figurative language versions of the models.
The results show that the use of contrastive learning with figurative language improves performance on the emotion classification task for most emotion types. For the micro-averaged F1-score, across all emotions, the figurative language versions of the models achieve an improvement of 6.60\%, 6.66\%, 3.63\%, and 3.90\% for BERT, RoBERTa, ALBERT, and CodeBERT respectively.
This implies that adding figurative language to these models improves their capability to comprehend and interpret the subtleties that developers use in their communication.

In all four models, we see an increase in True Positives and a decrease in both False Negatives and False Positives. For instance, for the BERT model, across 6 emotions, micro-averaged recall increases by 5.58\%, and micro-averaged precision increases by 7.59\%. This indicates improved precision in predictions after applying contrastive learning.

When considering the average improvement in individual emotions across all models in Table~\ref{tab:classification_results_emotions}, we observe that \textit{`Joy'} has most improvement (7.67\%), followed by \textit{`Surprise'} (5.45\%). This correlates with the frequency of occurrence of these emotions in GitHub comments, e.g., Joy is much more commonly found than Fear.
As our figurative language dataset is randomly sampled, it is likely to contain figurative expressions closely related to the emotions that are more commonly observed in GitHub.
This result suggests that curating a larger and more diverse set of comments that include figurative language could lead to a stronger and more balanced performance improvement.

\begin{table}[tb]
\centering
\scriptsize
\caption{Evaluation of LLMs finetuned with figurative language on the Emotions Dataset (F1-score).}
\def\arraystretch{0.9}
\begin{tabular} {|l|c|c|c|c|c|c|c|}
\hline
    Model & Anger & Love & Fear & Joy & Sad. & Surp. & Mic.Avg.
    \\ \hline\hline
    BERT & 0.506 & \textbf{0.712} & 0.536 & 0.579 & 0.636 & 0.594 &  0.588 \\
    BERT-FL & \textbf{0.547} & 0.709 & \textbf{0.562} & \textbf{0.608} & \textbf{0.661} & \textbf{0.632} & \textbf{0.627}\\
    \tiny{+/-} & \tiny{+8.10\%} & \tiny{-0.42\%}	& \tiny{+4.85\%} & 	\tiny{+5.01\%} & \tiny{+3.93\%} & \tiny{+6.40\%} & \tiny{+6.60\%} \\
    \hline

    RoBERTa & 0.525 &	0.683 & 0.500 & 0.613 & \textbf{0.673} & 0.592 & 0.593 \\
    RoBERTa-FL & \textbf{0.551} & \textbf{0.733} & \textbf{0.545} & \textbf{0.667} & 0.667 & \textbf{0.617} & \textbf{0.632} \\
     \tiny{+/-}& \tiny{+4.95\%}	& \tiny{+6.82\%} & \tiny{+8.26\%} & \tiny{+8.10\%}	& \tiny{-0.90\%}	& \tiny{+4.05\%} & \tiny{+6.66\%} \\
    \hline

    ALBERT & \textbf{0.462} & 0.658 & 0.430 & 0.487 & \textbf{0.628} & 0.564 &  0.531 \\
    ALBERT-FL & 0.443 & \textbf{0.682} & \textbf{0.435} & \textbf{0.540} & 0.624 & \textbf{0.592} & \textbf{0.550} \\
    \tiny{+/-} & \tiny{-4.11\%}	&\tiny{+3.52\%}	&\tiny{+1.15\%}	&\tiny{+9.81\%}	&\tiny{-0.64\%}	&\tiny{+4.73\%} & \tiny{+3.63\%} \\
    \hline

    CodeBERT & 0.484 & 0.711 &	\textbf{0.507} & 0.558 & 0.575 & 0.576 & 0.561 \\
    CodeBERT-FL & \textbf{0.497} & \textbf{0.723} & 0.444 & \textbf{0.605} & \textbf{0.645} & \textbf{0.617} & \textbf{0.583} \\
    \tiny{+/-} &\tiny{+2.79\%}	&\tiny{+1.70\%}	&\tiny{-14.08\%}	&\tiny{+7.75\%}	&\tiny{+10.92\%}	&\tiny{+6.61\%} & \tiny{+3.90\%} \\
    \Xhline{2\arrayrulewidth}
     Avg. +/- &+2.93\%	&+2.90\% &+0.04\%	&+7.67\% &+3.33\%	&+5.45\% & +5.20\%\\
    \hline
\end{tabular}
\vspace{-6pt}
\label{tab:classification_results_emotions}
\end{table}

\smallskip
\noindent
\textul{Error Analysis of BERT-FL vs. BERT.}
To gain deeper insight into figurative language-based models' predictive accuracy relative to baseline models, we perform qualitative analysis. Our focus is solely on BERT and BERT-FL models' predictions. We examine two specific areas: 1) True Positives where BERT-FL is correct while baseline BERT is not, and 2) True Positives where baseline BERT is correct while BERT-FL is not.

Among the positive instances, BERT-FL correctly predicts 39 utterances that the baseline BERT model does not.
Consider the following sentence: \sent{Bah. Wasn't supposed to add anything -- it was a debugging leftover...}. In this case, BERT-FL correctly predicts \textit{`Anger'}, whereas BERT misclassifies it. Here, the word \figlang{leftover} is used metaphorically. Normally, \figlang{leftover} refers to \figlang{something that remains unused or unconsumed}, particularly in the context of food\footnote{\url{https://www.merriam-webster.com/dictionary/leftover}}.
However, in the given sentence, the word is used to imply that some code or modifications were unintentionally left behind or overlooked during the debugging process. The BERT-FL model likely captures the context more effectively. Another example, \sent{Oh nice!! I've seen that syntax floating around, wanting to try it for a while \textit{`raising-hands`}} - BERT-FL correctly classifies as \textit{`Joy'} which the BERT baseline model misclassifies. Here, the BERT-FL is likely able to capture that \figlang{floating around} is an idiom\footnote{\url{https://idioms.thefreedictionary.com/floating+around}} and interpret the meaning. In some instances, BERT-FL makes correct predictions by adopting a more conservative classification approach. For instance, BERT classifies the following sentence as \textit{`Anger'}: \sent{Please put this below line 5 (together with the other non-app imports) :pray:}. However, BERT-FL accurately predicts that it is not \textit{`Anger'}.

On the other hand, in 27 cases, BERT-FL makes wrong predictions where BERT does not.
Consider this utterance: \sent{I have currently no clue, but I'll have a look}, this sentence contains the idioms \figlang{have a clue}~\footnote{\url{https://idioms.thefreedictionary.com/have+a+clue}} and \figlang{have a look}\footnote{\url{https://www.thefreedictionary.com/have+a+look}}. The author of the comment likely was puzzled about some functions or errors. BERT identifies correctly as \textit{`Surprise'} but BERT-FL does not. Possibly, BERT-FL interpreted these idiomatic expressions more of a literal interpretation of the words. In some cases, BERT-FL just misclassifies without any involvement of any figurative expressions. For example, \sent{I guess my concern is that it sets a precedent where somebody could see it and think that it would be fine to use in `core'.} This expression express concern which is annotated as \textit{`Fear'}. This expression conveys concern, annotated as \textit{`Fear'}. BERT identifies it correctly, but BERT-FL does not.
It is possible that during the contrastive learning process, BERT-FL may lose some of the baseline BERT model's ability to capture nuanced emotional indicators in certain sentences accurately.
This suggests that while this approach improves the overall model performance but may introduce limitations or biases in some cases.

\begin{table}[tb]
\centering
\scriptsize
\caption{Evaluation of LLMs finetuned with figurative language on the Incivility Dataset (F1-score).}
\def\arraystretch{0.9}
\begin{tabular} { |l|c|c|c| }
\hline
    Model & Civil & Uncivil & Micro Average
    \\ \hline\hline
    BERT & 0.537 & 0.814 & 0.734 \\
    BERT-FL & \textbf{0.587} & \textbf{0.853} & \textbf{0.783} \\
    \tiny{+/-} & \tiny{+8.54\%}	& \tiny{+4.84\%} & \tiny{+6.67\%} \\
    \hline

    RoBERTa & 0.424 & 0.827 & 0.734 \\
    RoBERTa-FL & \textbf{0.535} & \textbf{0.847} & \textbf{0.769} \\
    \tiny{+/-} & \tiny{+20.73\%}	&\tiny{+2.33\%} &\tiny{+4.76\%} \\
    \hline

    ALBERT & 0.151 & 0.807 & 0.685 \\
    ALBERT-FL & \textbf{0.423} & \textbf{0.809} & \textbf{0.713} \\
    \tiny{+/-} & \tiny{+64.28\%}	& \tiny{+0.30\%} & \tiny{+4.08\%} \\
    \hline

    CodeBERT & 0.185 & 0.810 & 0.692 \\
    CodeBERT-FL & \textbf{0.431} & \textbf{0.833} & \textbf{0.741} \\
    \tiny{+/-} & \tiny{+57.01\%}	&\tiny{+2.74\%} & \tiny{+7.07\%} \\
    \Xhline{2\arrayrulewidth}

    Avg. +/- & +37.64\% & +2.55\% & +5.65\% \\
    \hline

\end{tabular}
\vspace{-6pt}
\label{tab:classification_results_incivility}
\end{table}

\smallskip
\noindent
\textbf{\textit{Incivility Classification}}: Table~\ref{tab:classification_results_incivility} presents the results of the \textit{incivility classification} task on Ferreira et al.'s incivility dataset~\cite{ferreira2022heated}, using the same four large language models (BERT, RoBERTa, ALBERT, and CodeBERT) with and without the contrastive learning approach.
The micro-averaged F1-scores indicate that the models perform better when applying the contrastive learning approach. Overall, the BERT, RoBERTa, ALBERT, and CodeBERT models have an average improvement of 6.67\%, 4.76\%, 4.08\%, and 7.07\% respectively, when the contrastive learning approach is applied.
Since the incivility dataset is small and imbalanced, the baseline models often struggle to classify the minor \textit{`Civil'} class, except for BERT.

We also observed a significant average improvement of 37.64\% across all models in the \textit{`Civil'} class, compared to a modest 2.55\% improvement in the \textit{`Uncivil'} class. This discrepancy likely arises because the figurative language dataset used for contrastive learning primarily consists of \textit{`Civil'} comments, which are much more common on GitHub than \textit{`Uncivil'} comments. Incorporating more figurative expressions from \textit{`Uncivil'} comments into the dataset could potentially enhance performance in this category as well.

It is important to note that the substantial improvements in identifying \textit{`Civil'} comments are largely attributable to ALBERT and CodeBERT, which showed improvements of 180\% and 133\%, respectively. These models started from a lower performance baseline, making such large gains more achievable compared to other models. However, BERT and RoBERTa also demonstrated stronger performance improvements in the \textit{`Civil'} class.

\subsection{RQ3: Can a better understanding of figurative language enhance software engineering automation where affect plays a role?}

To answer this RQ, we focus on a specific use case: automatic bug report priority detection, a major research area in software engineering~\cite{tian2013drone, wang2022clebpi, tian2015automated, umer2019cnn}, where previous research has highlighted the role of affect~\cite{umer2018emotion}.

\smallskip
\noindent
\textul{Dataset.} Bugzilla bug reports are widely used for priority detection~\cite{tian2013drone, wang2022clebpi, umer2018emotion}. The bug priority reports in Bugzilla are divided into 5 classes (i.e., P1 to P5, where P1 represents the highest priority while P5 represents the lowest priority).
Wang et al. collected 220k bug reports from Bugzilla~\cite{wang2022clebpi}. We sample 25\% of this dataset using stratified sampling across the 5 classes. We sample separately from the training and testing splits provided by the authors, which yielded a total of 49.6k bug reports. The distributions provided by the authors are: 1) training: P1 - 19.56\%, P2 - 18.45\%, P3 - 58.12\%, P4 - 1.66\%, and P5 - 2.21\%; and 2) testing: P1 - 19.21\%, P2 - 17.66\%, P3 - 59.5\%, P4 - 1.48\%, and P5 - 2.15\%.

\smallskip
\noindent
\textul{Procedure and Metrics.} We use the same four LLMs (BERT, RoBERTa, ALBERT, and CodeBERT) as baselines and follow the same approach for training and testing described in RQ2. We use F1-score as evaluation metric.

\smallskip
\noindent
\textul{Results and Discussion.} Table~\ref{tab:classification_bug_report_priority} shows the results of bug report priority prediction on the Bugzilla dataset. All four models made small improvements (1.96\%, 2.40\%, 3.71\%, and 1.61\% respectively) when fine-tuned with figurative languages.
On the other hand, the improvements across classes (P1-P5) varied. The change in the P5 class was minimal (0.27\%), and none of the models succeeded in recognizing any of the P4 instances. This is likely due to the fact that these two classes have the smallest amounts of data, comprising only 1.66\% for P4 and 2.21\% for P5 of the training data, respectively. Such findings suggest that fine-tuning with figurative language is not beneficial in cases of extreme data imbalance. For the average performance improvement across all models in the other three bug priority classes, we observe that P3 improved least (1.43\%) while P1 and P2 make more substantial gains of 4.37\% and 8.23\%. Umer et al.~\cite{umer2018emotion} noted that a substantial number of instances in the Bugzilla dataset are \textit{`Neutral'}, indicating that including figurative expressions from \textit{`Neutral'} utterances — which our dataset predominantly omits — could potentially yield additional benefits.

\begin{table}[tb]
\centering
\scriptsize
\def\arraystretch{0.9}
\caption{Evaluation  of LLMs finetuned with figurative language on the Bug Report Priority dataset (F1-score).}
\begin{tabular} { |l|c|c|c|c|c|c| }
\hline
    Model & P1 & P2 & P3 & P4 & P5  & Micro Average
    \\ \hline\hline
    BERT & 0.606 & 0.329 & 0.833 & 0.0 & 0.663 & 0.716 \\
    BERT-FL & \textbf{0.632} & \textbf{0.359} & \textbf{0.842} & 0.0 & \textbf{0.667} & \textbf{0.730} \\
    \tiny{+/-} & \tiny{+4.31\%} & \tiny{+9.14\%} & \tiny{+1.10\%} & \tiny{-} & \tiny{+0.52\%} & \tiny{+1.96\%} \\
    \hline

    RoBERTa & 0.61 & 0.293 & 0.827 & 0.0 & \textbf{0.677} & 0.707 \\
    RoBERTa-FL & \textbf{0.624} & \textbf{0.343} & \textbf{0.839} & 0.0 & 0.674 & \textbf{0.724} \\
    \tiny{+/-} & \tiny{+1.91\%}	&\tiny{17.24\%}	&\tiny{+1.39\%}	&\tiny{-} &\tiny{-0.51\%} & \tiny{+2.40\%} \\
    \hline

    ALBERT & 0.564 & 0.288 & 0.810 & 0.0 & 0.670 & 0.683 \\
    ALBERT-FL & \textbf{0.602} & \textbf{0.299} & \textbf{0.827} & 0.0 & \textbf{0.674} & \textbf{0.709} \\
    \tiny{+/-} & \tiny{+6.71\%} &\tiny{+3.88\%}	&\tiny{+2.14\%} & \tiny{-} &\tiny{+0.53\%} &\tiny{+3.71\%} \\
    \hline

    CodeBERT & 0.608 & 0.363 & 0.830 & 0.0 & 0.667  & 0.714 \\
    CodeBERT-FL & \textbf{0.636}	& \textbf{0.373}  & \textbf{0.839} & 0.0  & \textbf{0.670}  & \textbf{0.726} \\
    \tiny{+/-} & \tiny{+4.55\%}	&\tiny{2.64\%}	&\tiny{+1.08\%} &\tiny{-}	&\tiny{0.52\%} &\tiny{+1.61\%} \\

    \Xhline{2\arrayrulewidth}
    Avg. +/- & +4.37\%	& +8.23\%	& +1.43\% & - & +0.27\%  & +2.42\% \\
    \hline
\end{tabular}
\vspace{-2pt}
\label{tab:classification_bug_report_priority}
\end{table}

\smallskip
\noindent
\textul{Error Analysis of BERT-FL vs. BERT.} To get an understanding of where fine-tuned models are getting results correctly compared to baseline models, we look into 51 instances where BERT-FL makes the right predictions but BERT does not. We find that, indeed, some of these bug reports include metaphors and idioms.
For example, consider the following bug report description, which is at the P2 priority level: \sent{Deadlock when adding JSF framework I have experienced a deadlock while I was adding JSF framework to regular web project. [...]} Here \figlang{Deadlock} is a SE-specific figurative expression. The baseline model predicted P3, but BERT-FL made the correct prediction.
Another example \sent{Toot your own horn, put your name in the credits window The credits window is empty [...]}, annotated as P3. Here, \figlang{toot your own horn} is an idiom. BERT-FL correctly predicted but the baseline model did not.

However, there are also cases with figurative language where the fine-tuned model predicted incorrectly, while the baseline model was right.
For example, consider the following bug report \sent{offline task data is not retrieved on query [...] (i.e., fetch all things before hitting the road). [...]} Here, \figlang{hitting the road} is an idiom\footnote{\url{https://www.thefreedictionary.com/hitting+the+road}}.
The BERT-FL model predicts P1 when the original label is P3. It is possible that BERT-FL recognizes the idiom, prioritizes its figurative meaning, and predicts a higher class than the original label.

\subsection{Implications}\label{implications_se_fig}

There are a number of actionable implications to our study. Creating a glossary of common figurative language for a software project can be an invaluable tool for efficiently onboarding new developers~\cite{dominic2020onboarding}. It would help newcomers understand project-specific or domain-specific terms, which are essential for their quick integration. Minimizing the use of obscure jargon that may cause misunderstandings can enhance mutual understanding and collaboration among project participants~\cite{junior2022c2m}. Lastly, it's important to consider cultural differences~\cite{junior2022c2m} that may influence the interpretation of figurative language, as these nuances can significantly affect comprehension and communication within a diverse team.

Our study paves the way for several promising research directions in the realm of figurative language comprehension within software engineering: 1) Integrating figurative language into cutting-edge software engineering tools, such as CleBPI~\cite{wang2022clebpi}, could be achieved through innovative approaches like contrastive learning, self-supervised learning, or adversarial training; 2) Investigating the role of figurative language in specific scenarios, including toxic or uncivil comments, bug reports, and documentation, may yield insights into its effects on software development workflows; 3) Exploring the use of figurative language as a means for data augmentation presents an intriguing opportunity, building on established data augmentation techniques~\cite{imran2022data}; 4) Broadening the scope of analysis to encompass various forms of figurative language, such as similes, hyperbole, and personification, could enhance the depth of model training; 5) Extending our analysis to software engineering communication platforms beyond GitHub, including Stack Overflow, Gitter, JIRA, and app reviews, would offer a more holistic view of figurative language usage across different settings. Adapting Large Language Models (LLMs) for domain-specific figurative language has recently garnered interest in the NLP community~\cite{joseph2023newsmet, hilton2022metaphor, naseem2022robust, wijesiriwardene2023we}. Our work compliments this by adapting LLMs to the figurative language in software engineering.
 \section{RELATED WORK}
We describe the related work sourced from three different domains: figurative language analysis in the domain of Natural Language Processing (NLP), affect analysis in Software Engineering (SE), and bug report analysis in SE.

\smallskip
\noindent
\textbf{Figurative Language in NLP.} Figurative language has long been a topic of study in the field of NLP~\cite{stowe2022impli, chakrabarty2021figurative, fussell1998figurative, esmaeilzadeh2022text}. Research has explored its impact across various communication channels, including online reviews and social media~\cite{liu2020interaction, kronrod2013wii}. Social media platforms frequently employ figurative language to convey emotions, opinions, and feelings~\cite{recupero2019frame}.
Furthermore, there have been investigations that focus on the use of figurative language within specific domains~\cite{joseph2023newsmet, hilton2022metaphor, naseem2022robust}.
Specific forms of figurative language, such as metaphors, idioms, similes, sarcasm, and irony, have been studied in relation to tasks like offensive language detection~\cite{plaza2022integrating, weitzel2016comprehension}.

Scholars have categorized the detection of figurative language into two tasks: recognizing text containing figurative language and interpreting figurative expressions to identify their intended literal meaning~\cite{shutova2010models}. Recognizing figurative language presents challenges due to the multiple interpretations that expressions can have. To address these challenges, various methods have been proposed, such as word vectors, rule-based approaches, semantic patterns, and the application of LLMs~\cite{hao2010ironic, peng2017automatic, su2020deepmet, gamage2022bert}. Interpreting figurative language is a more complex task that requires a deeper understanding of the text's meaning~\cite{mcglone1996conceptual}. Previous approaches have utilized knowledge-based and corpus-based methods~\cite{veale2008fluid, martin2006corpus}. Researchers have leveraged LLMs to paraphrase figurative expressions tasks and have been successful in interpreting metaphors, idioms, hyperbole, irony, sarcasm, and similes~\cite{stowe2022impli, joseph2023newsmet, chakrabarty2021figurative, chakrabarty2022flute}. Some of the most common strategies applied to interpret figurative expressions using LLMs are zero-shot learning and fine-tuning using contrastive learning~\cite{liu2022testing}. Different from the prior work, we investigate the ability of LLMs to interpret figurative language in the context of SE communication.

\smallskip
\noindent
\textbf{Affect Analysis in SE.} Affective expressions in written text can be effectively analyzed by identifying linguistic cues that convey emotions, feelings, or attitudes~\cite{liu2020sentiment, munezero2014they}. The field of affect analysis in software engineering is rapidly growing, focusing on understanding how emotions, opinions, sentiment, toxicity, incivility, burnout, and offensive language impact software development activities~\cite{lin2022opinion, sarker2020benchmark, novielli2018benchmark, ferreira2022heated, Chatterjee2021AutomaticEO, Chatterjee2020Journal, Murgia2014DoDF, novielli2018gold, chen2021emoji, imran2022data, Sajadi2023, sentimoji, linsentiment, sentistrengthSE, ferreira2021shut, ferreira2022incivility}.

LLMs have emerged as powerful tools for affect analysis, making significant strides in the software engineering domain~\cite{sun2019fine, ke2021adapting, chiorrini2021emotion, acheampong2021transformer}. They have proven their mettle in sentiment analysis across various software-related artifacts and have even been instrumental in detecting incivility and toxicity~\cite{zhang2020sentiment, eeshita-sentiment, ferreira2022incivility, sarker2022identification}.

Despite the advancements made and the use of modern LLMs, some limitations need to be addressed, particularly in terms of generalizability.
SE-specific affect analysis tools trained on one communication forum may not perform well when applied to another forum due to differences in norms, conventions, and cultures that influence the expression of emotions and sentiments~\cite{novielli2020can, novielli2021assessment, novielli2018benchmark}. One major reason for this limitation is the tools' inability to recognize implicit emotions or sentiments, often inferred through context, tone, or other cues~\cite{imran2022data, novielli2018benchmark}. These challenges call for developing more versatile and adaptable tools that can be applied across multiple domains. In this paper, we explore interpreting and fine-tuning figurative languages with LLMs to enhance the generalizability of SE-specific affect analysis tools across different artifacts.

\smallskip
\noindent
\textbf{Bug Report Analysis in SE.} Bug report analysis is a mature research area in SE spanning tasks like duplicate bug detection, bug localization, deficient bug report, bug severity prediction, and priority assignment~\cite{chaparro2019reformulating, ciborowska2022fast, chaparro2017detecting, imran2021automatically, gomes2019bug, tian2013drone, wang2022clebpi, umer2019cnn}. Of particular relevance is bug report priority prediction, where affects in report descriptions can influence triage decisions~\cite{umer2018emotion}. Recently, priority inference models based on deep learning have been proposed using LLMs~\cite{wang2022clebpi}. This study explores whether fine-tuning LLMs with figurative language can enhance performance of the task or not.
 \section{THREATS TO VALIDITY} Several limitations may impact the interpretation of our findings. We categorize and list each of them below.

\smallskip
\noindent{\textul{Construct validity.}} Construct validity refers to the degree to which the study measures the concepts and constructs it claims to measure.
A threat may arise from the manual annotations for the dataset, specifically in creating semantically similar EMS and DMS sentences. To mitigate this, we provided clear instructions and examples to the annotators.
Additionally, we only examined metaphors and idioms; including other figurative language may alter results. To investigate this, our annotation approach can be expanded to analyze other forms.
Another potential threat is that our figurative language dataset was sourced from developer communication in 9 GitHub repositories, which may not be representative of the figurative language present on GitHub.

\smallskip
\noindent{\textul{Internal validity.}} Internal validity concerns the extent to which the study's findings can be attributed to the manipulation of the independent variable.
A threat is that the improved affect analysis performance with figurative language fine-tuning may not be solely due to the figurative language. However, we see consistent improvements across all models and datasets, indicating it is a key factor.
Not doing cross-validation on the smaller datasets can be another threat. To mitigate this, we use stratified sampling for representativeness and a standard 80-20\% train-test split.

\smallskip
\noindent{\textul{External validity.}} External validity pertains to the generalization of the findings of our study to other settings and contexts.
Our results may not generalize beyond the specific studied models, datasets, and any other domain than GitHub. However, we use diverse pre-trained LLMs and a Bugzilla dataset, showing some cross-domain applicability. Further investigation is needed to validate our results beyond the tools, data, and platforms used in our study.

\section{CONCLUSION}
This paper examined the relevance and impact of figurative language in software engineering communication. To investigate this, we annotated metaphors and idioms in a set of 2000 sentences collected from GitHub issues and PRs which resulted in 1661 sentences with figurative expressions, conducted a comprehensive analysis of the prevalence of figurative language in messages posted on PRs and issues in top 100 GitHub repositories, fine-tuned several state-of-the-art pre-trained LLMs with the annotated dataset, and evaluated the performance of these fine-tuned models on three publicly available SE-specific datasets. Our results indicated that figurative language is prevalent in software engineering communication, and fine-tuning LLMs with figurative language leads to improved performance on affect analysis tasks (on the best model, 6.66\% improvement on a GitHub emotion dataset, 7.07\% improvement on a GitHub incivility dataset, and 3.71\% improvement on a bug report prioritization dataset). Overall, our findings provide evidence for the relevance and impact of figurative language in software engineering communication and the potential benefits of fine-tuning LLMs with figurative language in the context of software engineering. However, there is room for further investigation.

Beyond the future work directions discussed in Section~\ref{implications_se_fig}, our error analysis shows that fine-tuned models may sometimes overemphasize figurative language, motivating the need for a different fine-tuning approach. Addressing this issue while preserving interpretive abilities presents an area for future research. Experimenting with generative language models like ChatGPT and LLaMa to assess their potential in enhancing the automatic interpretation of complex figurative expressions could significantly benefit communication and understanding in software development contexts.
Overall, this study provides a starting point for further empirical research on figurative language’s impact on software engineering communications in different application domains.

\bibliographystyle{IEEEtran}
\bibliography{references}

\end{sloppypar}

\end{document}